\documentclass[14pt,a4paper]{article}
\usepackage[14pt]{extsizes}
\usepackage[utf8]{inputenc}
\usepackage[english]{babel}

\usepackage[pdftex,colorlinks,linkcolor=blue,
citecolor=blue]{hyperref}
\usepackage{amsmath, amssymb, amsthm}

\newcommand{\ket}[1] {\!\!\;\ensuremath{\left|#1\right\rangle}}

\newcommand{\ketbra}[2]{\!\!\:\ensuremath {\left|#1\right\rangle\!\:\!\!\left\langle#2\right|}}

\title{Computation of operator exponentials\\ using the Dunford-Cauchy integral}
\author{Alexander Tsirulev\\
\small \textit{Faculty of Mathematics, Tver State University, Tver, Russia, 170002}}
\date{}

\begin{document}

\maketitle

\begin{abstract}
We consider an $n$-qubit quantum system with a Hamiltonian, defined by an expansion in the Pauli basis, and propose a new algorithm for classical computing the exponential of the Hamiltonian. The algorithm is based on the representation of the exponential by the Dunford-Cauchy integral, followed by an efficient computation of the resolvent, and is suitable for Hamiltonians that are sparse in the Pauli basis. The practical efficiency of the algorithm is demonstrated by two illustrative examples.
\end{abstract}

\section{Introduction}\label{Sec1}

Functions of operators arise naturally as formal solutions to a number of problems in physics and applied mathematics and have been widely studied earlier in the equivalent formulation of functions of matrices~\cite{Moler2003, Higham2008}. The operator formulation is commonly used in simulating quantum Hamiltonian dynamics, quantum machine learning, quantum computing, etc. when a real quantum system is modeled by an artificial system of qubits with a suitable Hamiltonian and then the corresponding unitary transformation is approximately realized at the hardware level by a quantum circuit~\cite{Feynman1982, Lloyd1996,Kawase2023}. A preliminary classical calculation of the Hamiltonian exponential is useful both for testing quantum algorithms and for the inverse problem of choosing the model Hamiltonians itself; however, the computational complexity is exponential in the number of qubits. Moreover, we have no universal algorithm that would be computationally efficient for all types of Hamiltonians at once. We have Trotter-Suzuki decomposition~\cite{Suzuki1991}, Magnus expansion~\cite{Blanes2010}, and --- see review~\cite{Moler2003} --- Pad\'{e} approximation, truncated Taylor series (results will be not unitary if the series is not summed exactly), Jordan–Schur decomposition (not suitable for time-dependent Hamiltonians), etc.

In this paper, we propose a new algorithm for computing the exponential of a linear operator in a finite dimensional Hilbert space. The operator is assumed to be given by a sparse expansion in the Pauli basis, and its exponential is represented as the Dunford-Cauchy integral. An exact mathematical formulation of the problem is given in {Sec.\,\ref{Sec2}}. The algorithm is described in {Sec.\,\ref{Sec3}}. In {Sec.\,\ref{Sec4}}, we consider two illustrative examples. Conclusions and some prospects for quantum realization of the algorithm are presented in {Sec.\,\ref{Conclusions}}.

\section{Formulation of the problem}
\label{Sec2}

\vspace{-1ex}
Let $\mathcal{H}_n= \mathcal{H}^{\otimes n}$ be the Hilbert space of a quantum system consisting of $n$ qubits, and ${L(\mathcal{H}_n)}$ be the corresponding algebra of linear operators, so that ${\dim_{\:\!\mathbb{C}}\! \mathcal{H}_n=N}$ and ${\dim_{\:\!\mathbb{C}}\!L(\mathcal{H}_n)= N^2}$, where ${N=2^n}$. We consider the problem of efficient computing exponentials of the form  $\mathrm{e}^{-\beta\hat{H}}$, where $\hat{H}\in L(\mathcal{H}_n)$ is an arbitrary Hermitian operator, which will be regarded as a Hamiltonian. There are two special cases of interest in quantum theory. If $\beta=it$ or, in a more general context, when the Hamiltonian $\hat{H}$ is time-dependent and $\beta=i$, the exponential $\mathrm{e}^{-it\hat{H}}$ or, respectively, $\mathrm{e}^{-i\hat{H}(t)}$ describes the unitary dynamics of the corresponding quantum system. In another case, $\hat{H}$ denotes the Hamiltonian of a "small" subsystem weakly coupled to its thermal environment, and $\beta$ denotes the inverse temperature. Then the subsystem is described by the Gibbs state $\hat{\rho}=\mathrm{e}^{-\beta\hat{H}}/Z$ with the partition function $\mathcal{Z}=\mathrm{tr}\,\mathrm{e}^{-\beta\hat{H}}$.

Let $\{\hat{\sigma}_k\}_{k=0}^3$ (the identity is included) be the 1-qubit Pauli basis in ${L(\mathcal{H})}$. The Pauli basis in ${L(\mathcal{H}_n)}$ consists of the $n$-qubit Pauli operators
\begin{equation*}\label{}
\hat{\sigma}_K=\hat{\sigma}_{k_1\ldots{}k_n}= \hat{\sigma}_{k_1}\!\otimes\ldots\otimes\hat{\sigma}_{k_n}, \quad k_1,\ldots,k_n\!\in\!\{0,1,2,3\}.
\end{equation*}
Here and below, $K$ denotes both the string $k_1\ldots{}k_n$ and, alternatively, its decimal representation. Recall that the pairwise orthogonal operators $\hat{\sigma}_K$ are  Hermitian and unitary at the same time and satisfy the relations
\begin{equation}\label{tr-sigma}
\mathrm{tr}\,\hat{\sigma}_{0}= N, \;\,
\mathrm{tr}\,\hat{\sigma}_{K} \big|_{K\neq0}\!= 0, \;\,
\hat{\sigma}_{K}^2\!= \hat{\sigma}_{0}, \;\,
\hat{\sigma}_{K}\hat{\sigma}_{L}\!= S(K,L)\:\!\hat{\sigma}_{M},\;\, M\!=\!K\!\!*\!L,
\end{equation}
where ${0\leq{K,L}\leq N^2-1}$, ${S(K,L)\!\in\!\{\pm1,\pm i\}}$, and $*$ is the composition of Pauli strings (e.g., $312*210=102$). The procedures for calculating ${S(K,L)}$ and ${K\!\!*\!L}$ are quite simple, and we note only that $\hat{\sigma}_K$ and $\hat{\sigma}_L$ either commute or anticommute. Finally recall that the anti-Hermitian operators ${i\hat{\sigma}_K}$ form a basis in the real Lie algebra $\mathfrak{u}(N)$ of the unitary group $U(N)$ (and in the algebra $\mathfrak{su}(N)$ of the group $SU(N)$ if $K$ is restricted to the set ${\{1,\ldots,N^2\!-\!1\}}$). Indeed, it is easy to see that $[i\hat{\sigma}_K,i\hat{\sigma}_L]= C_{KL}i\hat{\sigma}_M$, where the structure constants $C_{KL}=iS(K,L)-iS(L,K)$ are real.

The computational algorithm under consideration is based on the representation of a Hamiltonian ${\hat{H}\in L(\mathcal{H}_n)}$ in the form
\begin{equation}\label{H-sigma}
\hat{H}= \sum\limits_{K\in\mathcal{T}}h_K\hat{\sigma}_{K},\quad h_K\in\mathbb{R},\quad \mathcal{T} \!\subset\!\{1,2,\ldots,N^2-1\},
\end{equation}
where we assume, without loss of generality, that the Hamiltonian is traceless. This Hamiltonian is also assumed \textit{to be \textbf{sparse}} (\textit{in the Pauli basis}), that is, the following two conditions are fulfilled: first, the condition $|\mathcal{T}|\ll N^2$ is satisfied, and second, the set ${\sigma_{\hat{H}}=\big\{\hat{\sigma}_K\big\}_{K\in\mathcal{T}} \bigcup\{\hat{\sigma}_0\}}$ in the expansion~(\ref{H-sigma}) is closed under composition; of course, some coefficients $h_K$ may be equal to zero. The first condition should be understood only as the necessity to relate the length of this expansion (including terms with zero coefficients) to available computational resources. The second is also not too restrictive, since the set $\sigma_{\hat{H}}$ can usually be extended to be closed (by successively adding to this set all operators of the form
${\hat{\sigma}_{K}\hat{\sigma}_{L}/S(K,L)\big|_{K,L\in\mathcal{T}}}$ that are not contained in it) while maintaining the first condition. As an example of a sparse operator, one can consider the expansion for the density operator of the so-called uniform quantum superposition,
\begin{equation*}
\hat{\rho}_s=
\frac{1}{N}\sum\limits_{k,l=0}^{N-1}\ketbra{k}{l}= \frac{1}{N}\sum\limits_{K\in\{0,1\}^n} \hat{\sigma}_K.
\end{equation*}
This expansion has $N^2$ terms in the computational basis, and only $N$ terms (which are closed under composition) in the corresponding Pauli basis. Hence, in certain cases, $\hat{\rho}_s$ can be considered as a sparse operator. On the other hand, the Hamiltonian of the quantum Heisenberg $XY$ model with a transverse magnetic field is obviously not sparse, since we will have $|\mathcal{T}|= N^2$ as a result of this procedure.

The Dunford-Cauchy integral for a Hamiltonian $\hat{H}$ is given by~\cite{Kato1995}
\begin{equation}\label{D-C}
\mathrm{e}^{-\beta\hat{H}}= \frac{1}{2\pi{}i} \int_{\gamma}
\mathrm{e}^{-\beta{z}}(z\hat{I}-\hat{H})^{-1}dz,
\end{equation}
where the contour $\gamma$ encloses the spectrum ${Sp(\hat{H})}$ of $\hat{H}$; the spectrum lies on the real axis and consists exactly of singular points of the resolvent ${(z\hat{I}-\hat{H})^{-1}}$. \textit{\textbf{Thus, the problem is to compute the integral}}~(\ref{D-C}) \textit{\textbf{for a Hamiltonian} $\hat{H}$ \textbf{of the form}}~(\ref{H-sigma}) \textit{\textbf{assuming that} $\hat{H}$ \textbf{is sparse}}.

Note that using the relations~(\ref{tr-sigma}) to sum the Taylor series, one obtains the expression $\exp(-\beta\hat{\sigma}_K)= \cosh\!\!\:\beta\,\hat{\sigma}_0- \sinh\!\!\:\beta\,\hat{\sigma}_K$ for the local Hamiltonian $\hat{\sigma}_K$. Exact summation of the Taylor series is possible in a more general case, which seems to be not reflected in the literature: \textit{if all the Pauli operators $\hat{\sigma}_{K}$ in the expansion}~(\ref{H-sigma}) (\textit{where $h_K$ may be time-depended}) \textit{anticommute pairwise, ${\{\hat{\sigma}_K,\hat{\sigma}_L\}= 2\delta_{K,L}\hat{\sigma}_0}$, then}
\begin{equation}\label{AntiComm}
\mathrm{e}^{-\beta\hat{H}}= \cosh(h\beta)\,\hat{\sigma}_0- \frac{\sinh(h\beta)}{h}\,\hat{H},
\quad
h=\left(\sum\limits_{K\in\mathcal{T}}h_K^2\right)^{\!1/2}.
\end{equation}
To prove this, it is enough to note that $\hat{H}^2=h^2\hat{\sigma}_0$.

\section{Algorithm}
\label{Sec3}

In this section, we briefly describe the computational scheme of the algorithm, omitting the discussion of some subtleties that may arise in its numerical realization. Keeping in mind the expansion~(\ref{H-sigma}), we represent the resolvent in the form (now $\hat{\sigma}_0$ have to be included in the Pauli expansion)
\begin{equation}\label{resolvent}
\big(z\hat{\sigma}_{0}- \hat{H}\big)^{-1}= r_0\hat{\sigma}_0+ \sum\limits_{K\in\mathcal{T}}r_K\hat{\sigma}_{K}.
\end{equation}
Using the identity
$\big(z\hat{\sigma}_{0}-\hat{H}\big)^{-1} \big(z\hat{\sigma}_{0}-\hat{H}\big)= \hat{\sigma}_{0}$ and the relations~(\ref{tr-sigma}),
one can write the system of linear equations for the coefficients $r_K$ as
\begin{multline}\label{sigma0=}
\!\!\!\left(r_0\hat{\sigma}_0+\! \sum_{K\in\mathcal{T}}r_K\hat{\sigma}_{K}\right)\!\! \left(z\hat{\sigma}_0-\! \sum_{L\in\mathcal{T}}h_L\hat{\sigma}_{L}\right)
=\hat{\sigma}_0zr_0-\! \sum_{L\in\mathcal{T}}\!\hat{\sigma}_{L}h_Lr_0\\
\vphantom{\int\limits_A^A}
+\!\sum_{K\in\mathcal{T}}\!\hat{\sigma}_{K}zr_K- \hat{\sigma}_{0}\!\sum_{K\in\mathcal{T}}\!h_Kr_K-\! \sum_{M\in\mathcal{T}}\!\hat{\sigma}_{M}\! \sum_{\mathcal{T}\ni{}K\neq{}M}\!a_{MK}r_K=\hat{\sigma}_0,
\end{multline}
where
\begin{equation}\label{a-MK}
a_{MK}= \sum\limits_{L\in\mathcal{T},\, K*L=M}\!\!h_LS(K,L)= \bar{a}_{KM},\quad K,M\in\mathcal{T}.
\end{equation}
Let us assume for the moment that $\mathcal{T}$ is an ordered set, $\mathcal{T}=\{K_1,\ldots,K_\tau\}$. Then, by redefining the indexation in~(\ref{H-sigma}) and (\ref{resolvent}) --- (\ref{a-MK}) by the rules $h_{K_i}=h_{i}$, $r_{K_i}=r_{i}$, and $a  _{K_iK_j}=a_{ij}$, one can write the system in the explicit form as

\begin{equation}\label{matrix-system}
\left(
    \begin{array}{ccccc}
      z & -h_1 & -h_2 & \dots  & -h_\tau \\
      -h_1 & z & -a_{12} & \dots  & -a_{1\tau} \\
      -h_2 & -a_{21} & z & \dots  & -a_{2\tau} \\
      \vdots & \vdots & \vdots & \ddots & \vdots \\
      -h_\tau & -a_{\tau1} & -a_{\tau2} & \dots  & z
    \end{array}
\right)\!\!
\left(
                                \begin{array}{c}
      r_0 \\ r_1 \\ r_2 \\ \vdots \\ r_\tau
    \end{array}
\right)=
\left(
    \begin{array}{c}1\\0\\0\\\vdots\\0\end{array}
\right).
\end{equation}

The main feature of the algorithm is the reduction of dimensionality of the problem: \textit{the matrix of the system}~(\ref{matrix-system}) \textit{is of size ${(1+\tau)\times(1+\tau)}$, where $\tau=|\mathcal{T}|\ll{N}$ for a sparse Hamiltonian}. This matrix has the form $zI-A$,\linebreak where ${A_{00}=0}$, ${A_{i0}=A_{0i}=h_K}$, ${A_{ij}=a_{ij}}$ if ${i\neq j}$, and ${A_{ii}=0}$, ${1\leq{i,j}\leq\tau}$. Solving equations~(\ref{matrix-system}), we obtain the resolvent coefficients, ${r_i(z,h),\, 0\leq{i}\leq\tau}$, as functions of $z$ and the coefficients ${h=(h_1,\ldots,h_\tau)}$. It should be emphasized once more that the applicability of the algorithm is restricted to Hamiltonians sparse in the Pauli basis. In other words, the set ${\big\{ i\hat{\sigma}_K \big\}_{K\in \mathcal{T}}}$ in~(\ref{resolvent}) (closed with respect to the commutator) spans some \textit{low-dimensional} subalgebra in $\mathfrak{su}(N)$. Below, without loss of generality, we will consider $\hat{H}$ to be restricted to $\mathfrak{su}(N)$, since the contribution of the identity operator to the exponential is simply a phase factor.

Since the spectrum ${Sp(\hat{H})}$ coincides with the set of zeros of the determinant  ${\det(zI-A)}$, it is naturally to compute the integral~(\ref{D-C}) using the residue theorem. Returning to the previous notations, we can write the final computational formula as
\begin{multline}\label{exp}
\mathrm{e}^{-\beta\hat{H}}= \sum\limits_{z\in Sp(\hat{H})} \mathrm{res}\big[\mathrm{e}^{-\beta{z}} (z\hat{\sigma}_{0}- \hat{H})^{\!-1}\big]\\
=\sum\limits_{K\in\mathcal{T}\bigcup\{0\}} \left\{\sum\limits_{z\in Sp(\hat{H})} \mathrm{res}\big[\mathrm{e}^{-\beta{z}} r_K(z,h)\big]\right\}\!\hat{\sigma}_{K}.
\end{multline}
In order to use this formula, we actually need only the first column of the inverse matrix $(zI-A)^{\!-1}$ as it follows from the right-hand side of~(\ref{matrix-system}).

\section{Two examples}
\label{Sec4}

In the first example, the simple, maybe time-dependent, Hamiltonian
\begin{equation}\label{H1}
\hat{H}_1= a\hat{\sigma}_{123}+ b\hat{\sigma}_{231}+ c\hat{\sigma}_{312}
\end{equation}
is sparse (if one agrees that ${3\ll 64}$), since ${\hat{\sigma}_{123}\hat{\sigma}_{312}= i\hat{\sigma}_{231}}$, ${\hat{\sigma}_{231}\hat{\sigma}_{123}= i\hat{\sigma}_{312}}$, ${\hat{\sigma}_{312}\hat{\sigma}_{231}= i\hat{\sigma}_{123}}$;
in the computational basis ${\left\{\ketbra{ijk}{pqr}\right\}}$, $\hat{H}_1$ contains 24 terms. For the resolvent
\begin{equation*}
(z\hat{\sigma}_{000}-\hat{H})^{-1}= \chi\hat{\sigma}_{000}+ \alpha\hat{\sigma}_{123}+ \beta\hat{\sigma}_{231}+ \gamma\hat{\sigma}_{312},
\end{equation*}
we obtain the system~(\ref{matrix-system}) in the form
$$
\begin{array}{rcl}
z\chi-a\alpha-\,b\beta\,-\,c\gamma\, &\!\!=\!\!& 1,\\
-a\chi+z\alpha+ic\beta-ib\gamma &\!\!=\!\!& 0,\\
-b\chi-ic\alpha+z\beta+ia\gamma &\!\!=\!\!& 0,\\
-c\chi+ib\alpha-ia\beta+z\gamma &\!\!=\!\!& 0,
\end{array}
$$
and its solution is
$$
\chi\!=\!\dfrac{z}{z^2\!-p^2},\quad \alpha\!=\!\dfrac{a}{z^2\!-p^2},\quad \beta\!=\!\dfrac{b\vphantom{\int\limits^a}}{z^2\!-p^2},\quad \gamma\!=\!\dfrac{c}{z^2\!-p^2},
$$
where $p^2\!=\!a^2\!+\!b^2\!+\!c^2$. Thus, we obtain
$$
\mathrm{e}^{-it\hat{H}}= \cos(pt)\,\hat{\sigma}_{000}- i\:\!(\sin(pt)/p)\,\hat{H}_1
$$
by summing in~(\ref{exp}) over two simple poles $z=\pm p\:\!$. In the Hamiltonian~(\ref{H1}), the Pauli operators  anticommute, so that this result is in accordance with the general formula~(\ref{AntiComm}).

The second example, more complicated but also purely illustrative, is chosen so that it can be treated analytically using Maple or Mathematica. We consider a cluster of four qubits with the sparse Hamiltonian \vspace{-1.0ex}
\begin{equation*}\label{}
\hat{H}_2= h_1\hat{\sigma}_{0123}+ h_2\hat{\sigma}_{0213}+ h_3\hat{\sigma}_{0330}+ h_4\hat{\sigma}_{1023}%\\multline
+h_5\hat{\sigma}_{1100}+ h_6\hat{\sigma}_{1230}+ h_7\hat{\sigma}_{1313},
\end{equation*}
which contains 64 terms in the computational basis. The matrix ${zI\!-\!A}$ in~(\ref{matrix-system}), its determinant and the cofactor of its first diagonal element have the form
\begin{equation*}
zI-A=\left(
\begin{array}{rrrrrrrr}
z\,\,& -h_1 & -h_2 & -h_3 & -h_4 & -h_5 & -h_6 & -h_7 \\
-h_1 &z\,\, & -h_3 & -h_2 & -h_5 & -h_4 &  h_7 &  h_6 \\
-h_2 & -h_3 & z\,\,& -h_1 & ih_6 &-ih_7 &-ih_4 & ih_5 \\
-h_3 & -h_2 & -h_1 &z\,\, &-ih_7 & ih_6 &-ih_5 & ih_4 \\
-h_4 & -h_5 &-ih_6 & ih_7 &z\,\, & -h_1 & ih_2 &-ih_3 \\
-h_5 & -h_4 & ih_7 &-ih_6 & -h_1 &z\,\, & ih_3 &-ih_2 \\
-h_6 &  h_7 & ih_4 & ih_5 &-ih_2 &-ih_3 &z\,\, &  h_1 \\
-h_7 &  h_6 &-ih_5 &-ih_4 & ih_3 & ih_2 &  h_1 &z\,\,
\end{array}
\right)\!,
\end{equation*}
\begin{equation}\label{det}
\det(zI-A)= \big[(z+h_1)^2-\mu+\nu\big]^2 \big[(z-h_1)^2-\mu-\nu\big]^2,
\end{equation}\vspace{-2ex}
\begin{equation}\label{cofactor}
\mathrm{Ad}^{(00)}_{zI-A}= \big[(z+h_1)^2-\mu+\nu\big] \big[(z-h_1)^2-\mu-\nu\big]\big[z^3 - (h_1^2+\mu)z - h_1\nu\big],
\end{equation}
where\vspace{-2ex}
$$
\mu = h_2^2+h_3^2+h_4^2+h_5^2+h_6^2+h_7^2,\; \nu = 2h_2h_3+2h_4h_5-2h_6h_7.
$$
If the cluster is in a thermal environment, it is easy to find its density operator in an analytical form. For brevity, we find only the partition function, which is determined by the coefficient ${r_0(z,h)}$ in~(\ref{resolvent}), that is, by the first diagonal term of the inverse matrix ${(zI-A)^{\!-1}}$. The direct calculation gives \vspace{-1.0ex}
\begin{equation}\label{r0}
r_0=\frac{\mathrm{Ad}^{(00)}_{zI-A}}{\det(zI-A)}= \frac{z^3 - (h_1^2+\mu)z - h_1\nu} {\big[(z+h_1)^2-\mu+\nu\big] \big[(z-h_1)^2-\mu-\nu\big]}\,.
\end{equation}
There are four simple poles (doubly degenerate eigenvalues of $\hat{H}_2$), namely,
\begin{equation*}
 z_1^\pm=-h_1\pm\sqrt{\mu-\nu},\quad  z_2^\pm=h_1\pm\sqrt{\mu+\nu}.
\end{equation*}
In accordance with the formula~(\ref{exp}), we obtain
\begin{multline*}
\mathcal{Z}= {\sum}_{z\in \{z_1^\pm,z_2^\pm\}} \mathrm{res}\big[\mathrm{e}^{-\beta{z}} r_0(z,h)\big]\\
\vphantom{\int\limits^A}
=\frac{1}{2}\mathrm{e}^{\beta h_1}\cosh(\beta\sqrt{\mu-\nu})   +\frac{1}{2}\mathrm{e}^{-\beta h_1}\cosh(\beta\sqrt{\mu+\nu}).
\end{multline*}
This partition function and, consequently, the thermodynamical properties of the cluster possess a very high symmetry; it is seen from the invariance of $\mathcal{Z}$ under the transformation $h_1\mapsto-h_1, \nu\mapsto-\nu$, and from the expressions $\mu\pm\nu=(h_2\pm h_3)^2+(h_4\pm h_5)^2+(h_6\mp h_7)^2$ whose symmetries are obvious. An interesting observation (as a result of computational experiments) consists in that the high symmetry of $\mathcal{Z}$ is normal for sparse Hamiltonians, that is, clusters with very different interactions between qubits may have the same thermodynamical properties.

These examples illustrate three important features of the algorithm. First, if the set ${\big\{ i\hat{\sigma}_K \big\}_{K\in \mathcal{T}}}$ in~(\ref{resolvent}) has minimally sufficient number of Pauli operators, then the matrix in~(\ref{matrix-system}) is non-singular everywhere except for the spectrum ${Sp(\hat{H})}$, despite the degeneracy of $\hat{H}$. Second, the algorithm presupposes that a given Hilbert space $\widetilde{\mathcal{H}}$ is of dimension $2^n$. Sometimes this restriction can be effectively overcome by embedding $\widetilde{\mathcal{H}}$ into $\mathcal{H}_n$, where $\dim\widetilde{\mathcal{H}}<2^n$. For example, let $\widetilde{\mathcal{H}}$ be the Hilbert space of a qutrit with the Hamiltonian ${\hat{H}'\!= 2\ketbra{\tilde{0}}{\tilde{0}}\!-\! 4i\ketbra{\tilde{1}}{\tilde{2}}\!+\! 4i\ketbra{\tilde{2}}{\tilde{1}}\!-\! 2\ketbra{\tilde{2}}{\tilde{2}}}$.
We define an embedding $\widetilde{\mathcal{H}}\!\mapsto\!\mathcal{H}_2$ using the following correspondence of their bases: ${\ket{\tilde{0}}}\mapsto\ket{00}$, ${\ket{\tilde{1}}}\mapsto\ket{01}$, ${\ket{\tilde{2}}}\mapsto\ket{10}$. The image of $\hat{H}'$ in $\mathcal{H}_2$ has the form $\hat{H}'= \hat{\sigma}_{30}+ \hat{\sigma}_{33}- 2\:\!\hat{\sigma}_{12}+ 2\:\!\hat{\sigma}_{21}$.
As another example, one can consider an analogous embedding of $\widetilde{\mathcal{H}}^{\otimes5}$ (of dimension 243) into $\mathcal{H}_8$. And third, the polynomials~(\ref{det}) and~(\ref{cofactor}) are not relatively  prime, and so the fraction~(\ref{r0}) has been reduced to lowest terms. Such a situation always occurs when the eigenvalues of the Hamiltonian are degenerate; this feature of the algorithm should be taken into account if the system~(\ref{matrix-system}) is solved numerically. Note also that the case of very closed eigenvalues requires a special approach using regularization techniques.\vspace{-2ex}

\section{Conclusions}
\label{Conclusions}\vspace{-1ex}

A new algorithm for classical computing Hamiltonian exponentials in the Pauli basis is proposed. The algorithm is based on the representation of the exponential by the Dunford-Cauchy integral and seems to be quite practical, at least for Hamiltonians that sparse in the Pauli basis. We have restricted our consideration to Hamiltonian exponentials only for the sake of definiteness, whereas the algorithm is applicable, with some minimal additions, to any linear operator in~${L(\mathcal{H}_n)}$ and any holomorphic function of the operator. Note that there are some attractive prospects for a quantum realization of the algorithm. Two direct approaches, probably not the most optimal, are quite obvious. First, the quantum algorithm of Ref.~\cite{Harrow2009} can be applied to solve the linear system~(\ref{matrix-system}), and second, in the spirit of Ref.~\cite{Takahira2020}, the calculation of residues by numerical integration can be performed on a quantum device.

\end{document}